\documentclass{article}
\usepackage{frascatiphys}
\usepackage{graphicx}
\begin{document}
\title{ 
THE GROWING TOOLBOX OF PERTURBATIVE QCD}
\author{
Lorenzo Magnea        \\
{\em Dipartimento di Fisica, Universit\`a di Torino, and INFN, Sezione di Torino} \\
}
\maketitle
\baselineskip=11.6pt
\begin{abstract}
Advances in perturbative QCD techniques have been crucial for the successful 
interpretation of the data collected in Run I of LHC, and for the discovery of the 
Higgs boson. I will very briefly highlight some recent additions to the QCD toolbox,
and note how these new tools are likely to be essential for future precision physics,
both in Run II at the LHC, and in view of future hadron and lepton colliders\footnote{Talk 
given at the Workshop ``LFC15: Physics Prospects for Linear and Other Future Colliders'',
ECT*, Trento, 7-11 September 2015.}.
\end{abstract}
\baselineskip=14pt

\section{Introduction}

The first run of the LHC was a resounding success, culminating in the Nobel-prize-winning
discovery of the Higgs boson: a great achievement, although the discovery was to a large 
extent expected. Strikingly, on the other hand, the Standard Model of particle physics held 
up, and it is now tested and verified to an unexpected, even amazing degree of accuracy
all the way up to the TeV energy scale: no evidence for new physics turned up in Run I. This 
seems to have heightened the expectations for Run II: indeed, the recent announcement by 
CMS and ATLAS of a small excess of events in the di-photon channel triggered the publication,
in less than one week, of more than one hundred papers with tentative theoretical interpretations, 
with the first papers appearing within minutes of the announcement. In a few months we will 
know if this outburst of speculative activity will be justified by further data. The task of this 
Workshop, however, is to look further ahead, to the next generation of machines which
are currently being discussed and planned, and which will succeed or complement the 
LHC at the high-energy frontier.

The lesson that I would like to draw from the experience of the past years, leading up 
to the LHC operation and the data analyses of Run I, is that the role of precision Standard
Model phenomenology has been crucial to develop a sufficient understanding of the
immensely complex processes underlying LHC collisions, and will remain crucial for
our ability to adequately exploit any future high-energy collider\cite{Forte:2015cia}. 

The past ten to fifteen years have seen remarkable progress in our quantitative control 
of the three stages of hadron collisions. The parametrisation of initial states by means 
of parton distributions (PDFs) has undergone a radical overhaul, and we now have 
several independent and reliable sets of PDF's, with credible determinations of their 
uncertainties\cite{Butterworth:2015oua}; our understanding of the hadron jets that 
characterise most collider final states has similarly evolved from qualitative to precisely 
quantitative, with the development of fast infrared-safe jet algorithms allowing for 
precise predictions for complex jet configurations, including studies of the internal structure 
of the jets themselves\cite{Adams:2015hiv}. Finally, our capabilities to compute the 
hard-scattering partonic cross sections at the heart of LHC collisions has progressed 
much beyond what might have been expected: NLO calculations of multi-particle final 
states matched to parton showers are now the standard, and the extension of these 
techniques to NNLO and beyond is well under way\cite{Andersen:2014efa}.

It is easy to argue that the splendid results of LHC Run I would not have been possible 
without this vast body of work, stemming from many collaborations involving hundreds
of phenomenologists. Similarly, exploiting future colliders, which will operate at even higher 
energies, and likely require even higher precisions, will not be possible without a continued 
effort to refine our understanding of Standard Model processes.

In the limited space of this contribution, I will begin by emphasising the non-trivial role 
played by QCD predictions even at future lepton colliders; I will continue by giving
some examples of the QCD tools developed in the past few years to handle high-order
perturbative calculations, and I will conclude by briefly summarising some recent 
progress in the field of soft-gluon resummation, which may soon shed light on a new
class of all-order contributions to interesting hadronic cross sections.

\section{QCD at future (lepton) colliders}

There is clearly no need to make the case for the importance of perturbative QCD
studies at future hadron colliders, such as foreseen upgrades of the LHC, or the
prospective 100 TeV collider\cite{Arkani-Hamed:2015vfh}. On the other hand, 
preliminary physics assessment of proposed lepton colliders, such as TESLA, 
ILC or CLIC, have often focused (quite understandably) on their new physics 
potential, leaving the Standard Model on the sidelines. On occasions, this 
emphasis can be misleading, and further analysis shows that a detailed 
high-precision Standard Model analysis is necessary in order to exploit 
the full potential of the machine. Here are a few examples, focusing on 
QCD studies.

\subsection{Hadronic jets}

Lepton colliders are designed as precision machines, but, at high energies, many
important final states will be characterised by a very high jet multiplicity. Such states
are not easy to characterise accurately. As an example, consider $t \bar{t} H$ 
production, with all particles decaying hadronically: this leads to an eight-jet final 
state, with at least four $b$-quark jets. If coloured supersymmetric particles were 
to be discovered, they would easily lead to even more complex final states. At a 
hadron collider, one might sidestep the problem by focusing on (semi) leptonic
final states, but given  the lower number of events to be expected for example at ILC,
exploiting fully hadronic final states may prove necessary. Such high-multiplicity 
final states are likely to require the most advanced available QCD 
techniques for jet identification, tagging and mass reconstruction. One may also 
note that some of these techniques will need to be retuned (see for example 
\cite{Boronat:2014hva}): boost invariance of the jet-finding algorithm will be less 
relevant, and jet-substructure studies will have a more limited impact since heavy 
states are unlikely to be heavily boosted.

\begin{figure}[htb]
    \begin{center}
        {\includegraphics[scale=0.12]{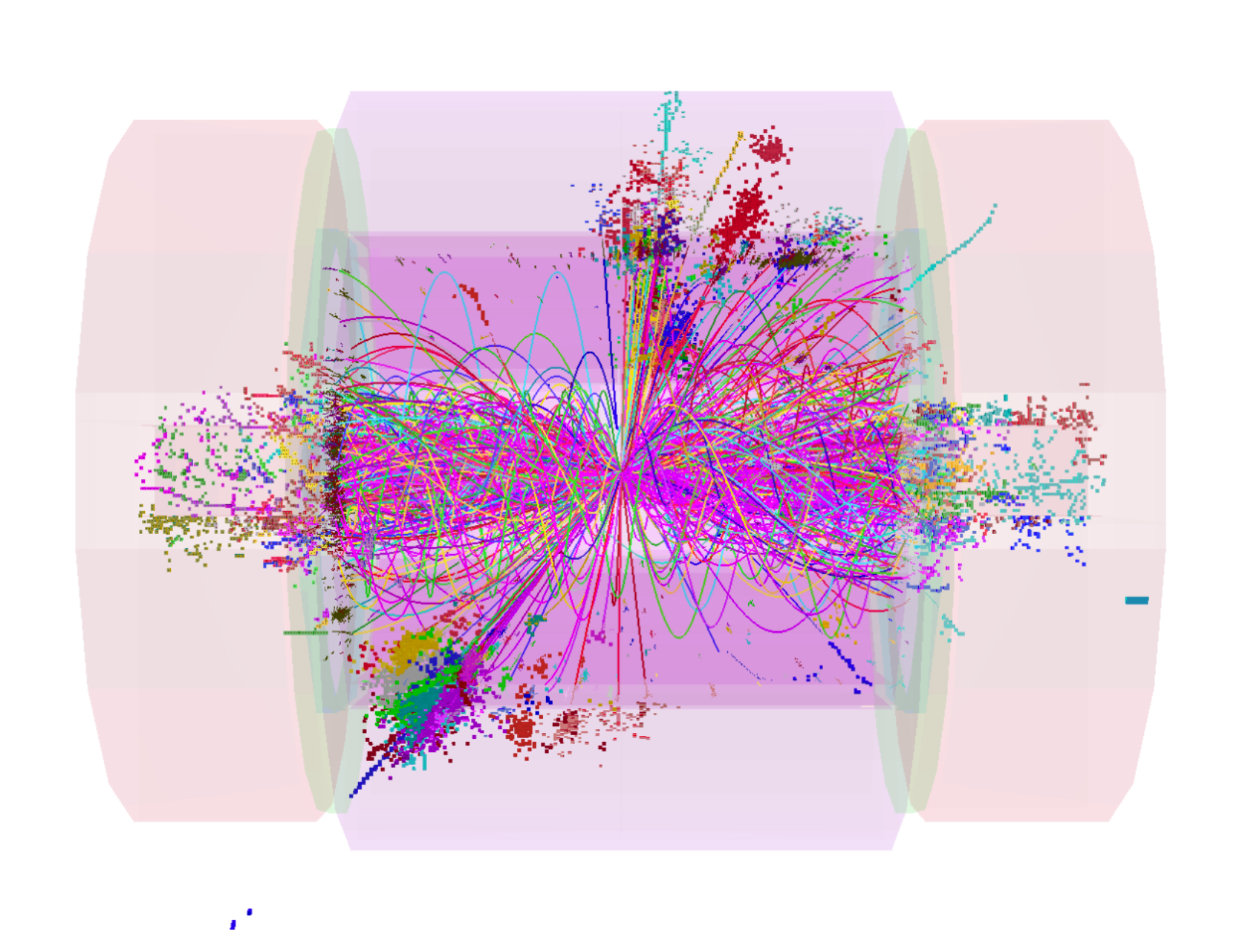}}\hspace{0.5cm}
        \caption{\it A simulated event including the production of a $t \bar{t}$ pair 
        at CLIC, with $\sqrt{s} = 3$ TeV, and overlaid background from $\gamma \gamma 
        \to$ hadrons, from\cite{Linssen:2012hp}.}
    \label{clic}
    \end{center}
\end{figure}

\subsection{Underlying event}

One of the reasons why lepton colliders are (correctly) touted as `clean' precision
machines is the absence of the `underlying event', the complex low $p_T$ scattering
of hadron remainders that surrounds the hard scattering at hadron colliders. It is however
well understood by now that at sufficiently high energy a very significant `underlying 
event' develops at lepton colliders as well. Just as protons at high energy can be 
seen as made mostly of gluons, leptons acquire an increasingly dominant photon 
component, which materialises as an underlying event through photon scattering,
via $\gamma \gamma \to$ hadrons. Fig.~\ref{clic}, taken from the CLIC Conceptual
Design Report\cite{Linssen:2012hp} shows the simulation of a hard scattering event 
including the production of a $t \bar{t}$ pair, at $\sqrt{s} = 3$ TeV, with the hadron 
background generated by photon collisions. At this CM energy, the background 
deposits 1.2 TeVs of energy per event in the detector, which will have to be subtracted
using refinements of recently developed tools such as jet areas\cite{Cacciari:2015jwa}.

\subsection{Standard Model parameters}

Lepton colliders hold the promise to give the most precise determinations of key 
Standard Model parameters, for example $m_{\rm top}$ and $\alpha_s$. This was 
discussed elsewhere in this Workshop, it has recently been reviewed in detail
in\cite{Moch:2014tta,d'Enterria:2015toz}, and certainly cannot be discussed in 
this very limited space. Once again, however, it is worth emphasising that these
determinations must rely upon state-of-the-art, high-order, precision QCD 
calculations. A case in point is the recently computed three-loop correction to
the near-threshold production of $t \bar{t}$ pairs\cite{Beneke:2015kwa}, which
will play a key role in the determination of $m_{\rm top}$ with better than permil 
precision through a threshold scan: only at this level, reached through a combination
of effective field theory techniques with high-level tools for loop calculations, one
observes that the theoretical uncertainty comes under full control.

\section{Selected examples of new tools}

Recent years have seen a remarkable degree of progress in our ability to compute
gauge theory amplitudes and cross sections to very high perturbative orders. To 
some extent, this was certainly triggered by the needs of LHC, but it is interesting
to note that several of the new techniques that have been deployed are connected 
to purely theoretical developments originating from studies of $N = 4$ Super-Yang-Mills 
(SYM) theory and thus ultimately related to string theory. Altogether, the new developments 
are feeding a `NNLO revolution' which has already yielded a number of phenomenologically
relevant results for $2 \to 2$ LHC processes. Some aspects of these recent developments 
are briefly touched upon below. 

\subsection{High-order amplitudes and iterated integrals}

The development of unitarity-based methods to compute scattering amplitudes, 
together with several pioneering high-order calculations in $N= 4$ SYM, brought 
the focus on the concept of `transcendental weight' of the functions arising in 
Feynman diagram calculations. We now know that a vast class of gauge-theory 
scattering amplitudes can be expressed in terms of generalised polylogarithms 
that can be generated by means of iterated integrals, which in turn encode in 
a simple way the singularity structure of the amplitude as a function of the 
Mandelstam invariants. Understanding the class of functions that make up the
result for a scattering amplitude can often turn an extremely difficult analytic 
problem into a relatively simple algebraic one, so these new mathematical tools
(recently reviewed in\cite{Duhr:2014woa}) have quickly found application in a 
number of phenomenological calculations. While the tools turn out to be especially
powerful for a conformal theory like $N=4$ SYM, it has become clear that they 
have direct applications also to QCD and electroweak amplitudes and cross 
sections. The breakthrough\cite{Henn:2013pwa} was the realisation that the well-known
method of differential equations for the computation of Feynman amplitudes could
be optimised to a truly remarkable degree by choosing (when possible) a basis of 
master integrals belonging to the class of iterated integrals mentioned above.
The method, reviewed in\cite{Henn:2014qga}, is proving very powerful, and the 
list of NNLO calculations that have become available in its wake is already much
too long to be referenced here. More generally, it is remarkable that, after many
decades of intensive studies, perturbative quantum field theory can still surprise 
us, with the discovery of new and beautiful mathematical structures and entirely
novel viewpoints.

\subsection{NNLO subtraction}

The calculation of loop-level partonic cross section requires the cancellation of infrared
and collinear divergences which appear separately in virtual corrections and when 
real emission corrections are integrated over the phase space of undetected partons.
The problem has been well understood in principle for decades, but the construction 
of a sufficiently general and efficient algorithm to perform the cancellation at NNLO
has proved much harder than expected. Crucially for phenomenological applications,
several practical solutions to this problem have now been proposed and are in different
stages of being applied or tested\cite{Currie:2012qxa,Czakon:2014oma,Bonciani:2015sha,
DelDuca:2015zqa,Boughezal:2015dra}. As a matter of principle, the optimal `subtraction
algorithm' should have several attributes: complete generality across all IR-safe 
observables with arbitrary numbers of final state partons, exact locality of the IR and
collinear counterterms, which should be computed analytically to optimize speed and
theoretical understanding, exact independence on external parameters introduced 
to `slice' away the singular regions of phase space, and overall computational efficiency.
In this sense, none of the existing methods qualifies as a `silver bullet' enjoying all 
these properties. The methods however have proven sufficiently powerful to perform
pioneering and highly non-trivial NNLO calculations, such as the $t \bar{t}$ 
production cross section\cite{Czakon:2013goa} and the Higgs-plus-jet cross 
section\cite{Boughezal:2015dra}. Rapid further developments towards the automatisation
of NNLO calculations, similarly to what has been done at NLO in recent years, are 
under way.

\subsection{Threshold resummation beyond leading power}

To conclude this bird's eye overview with a theme where I have made a direct contribution,
I will now briefly discuss the all-order summation of soft and collinear gluon effects, which
is often necessary to extend the applicability of perturbative calculations to regions of phase
space where large logarithms of ratios of mass scales appear order by order in the coupling.
Specifically, I consider the common situation in which a partonic cross section has a 
threshold for the production of some heavy state, for example a vector boson, a Higgs boson,
or a heavy coloured final state such as a $t \bar{t}$ pair. In these circumstances, the 
cross section $\sigma(\xi)$ depends logarithmically on the distance from threshold $\xi$,
according to
\begin{equation}
  \frac{d \sigma}{d \xi} \, = \, \sum_{n = 0}^{\infty} \left( \frac{\alpha_s}{\pi} \right)^n \, 
  \sum_{m = 0}^{2 n - 1} \left[ c_{n m}^{(-1)}
  \left( \frac{\log^m \xi}{\xi} \right)_+ + c_n^{(\delta)} \, \delta(\xi) +
  c_{nm}^{(0)} \, \log^m \xi \, + \, \ldots \right] \, .
\label{thresholddef}
\end{equation}
The leading-power logarithms determined by the coefficients $c_{n m}^{(-1)}$ are directly
related to the infrared and collinear divergences of the amplitudes, and, as a consequence,
they can be resummed to all-orders in perturbation theory, using a technology which has
been well understood for decades and is now routinely applied to increasing logarithmic 
accuracy. For massless gauge-theory scattering amplitudes, soft and collinear effects
factorise\cite{Dixon:2008gr}, according to
\begin{equation}
  {\cal A}_{n} (p_i) \, = \, \prod_{i = 1}^n \left[ \frac{J_i (p_i)}{{\cal J}_i (\beta_i)} \right] 
  \cdot {\cal S}_{n} (\beta_i) \cdot {\cal H}_{n} (p_i) \, ,
\label{factorised}
\end{equation}
where I wrote the particle momenta  as $p_i = Q \beta_i$, with $Q$ a hard scale, the soft 
function ${\cal S}_{n} (\beta_i)$ parametrises soft-gluon effects, and the jet functions $J$ 
and ${\cal J}$ contain collinear dynamics. Each function has a gauge invariant operator 
definition, for example for a quark
\begin{equation}
  J (p, n) \, u(p) \, = \, \left\langle 0 \left| \Phi_n (\infty, 0)
  \psi(0) \right| p \right\rangle \, ,
\label{Jdef}
\end{equation}
where $\Phi_n$ is a Wilson line factor and $n$ is an auxiliary `factorisation vector'. 
For well-behaved IR-safe observables, the factorisation in Eq. (\ref{factorised}) leads 
to resummation of leading-power threshold logarithms. At next-to-leading power (NLP), 
an increasing body of evidence has been suggesting that a similar organisation of the 
logarithms determined by the coefficients $c_{n m}^{(0)}$ should be possible\cite{Laenen:2008ux}. 
In the soft sector, it is indeed possible to extend the soft exponentiation theorem beyond 
leading power\cite{Laenen:2008gt,Laenen:2010uz}, but this proves insufficient
to generate all NLP logarithms starting at two loops. The reason is the interference
of collinear singularities with (next-to-) soft emissions, which prevents their complete 
factorisation. This obstacle was first overcome by Del Duca\cite{DelDuca:1990gz},
and recently revisited and applied to electroweak annihilation cross sections
in\cite{Bonocore:2014wua,Bonocore:2015esa}. The result is a generalisation of 
the leading-power factorisation in Eq. (\ref{factorised}), which, in its simplest form, 
reads
\begin{equation}
  {\cal A}^\mu (p_j, k) \, = \, \sum_{i = 1}^2 \left(q_i \, \frac{(2 p_i - k)^\mu}{2 p_i 
  \cdot k - k^2} + q_i \, G^{\nu\mu}_i \frac{\partial}{\partial p_i^\nu} + G^{\nu\mu}_i 
  J_\nu (p_i, k) \right) {\cal A} (p_i; p_j) \, ,
\label{NEfactor3}
\end{equation}
where ${\cal A}^\mu$ is an amplitude including the radiation of an extra soft gluon,
$G_{\mu \nu}$ is a kinematic projection, and $J_\mu$ is a `radiative jet' function 
defined by
\begin{equation}
  J_\mu \left( p, n, k \right) u(p) \, = \, 
  \int d^d y \,\, {\rm e}^{ - {\rm i} (p - k) \cdot y} \, \left\langle 0 \left| \,
   \, \Phi_{n} (y, \infty) \, \psi (y) \, j_\mu (0) \, \right| p \right\rangle \, ,
\label{Jmudef}
\end{equation}
where $j_\mu$ is the current for the production of the extra soft gluon.
Using Eq. (\ref{NEfactor3}), it is possible to exactly reproduce all NLP
logarithms at two loops for vector boson production cross sections, in
terms of universal soft and collinear factors. This strongly suggests that 
a complete resummation formalism for NLP logarithms is at hand, which
would then lead to a number of phenomenological applications to 
precision calculations of QCD cross sections of relevance for LHC 
and future colliders. Work is in progress to proceed in this direction.

\section{Acknowledgements}
Work supported by the Research Executive Agency (REA) of the European Union 
under the Grant Agreement PITN-GA-2012-316704 (HIGGSTOOLS); by MIUR (Italy), 
under contract 2010YJ2NYW$\_$006, and by the University of Torino and the 
Compagnia di San Paolo under contract ORTO11TPXK.

\end{document}